# Phylogenetic Study of 2019-nCoV by using alignment free method
# – Evolutionary Bifurcation of Novel Coronavirus Mutants

Yang Gao[1], Tao Li[2], and Liaofu Luo[‡3]

[1] Baotou National Rare Earth Hi-Tech Industrial Development Zone, Baotou, China

[2] College of Life Sciences, Inner Mongolia Agricultural University, Hohhot, China.

[3] School of Physical Science and Technology, Inner Mongolia University, Hohhot, China

[‡]Corresponding author.

E-mail: lolfcm@imu.edu.cn   (LL)

**Abstract**

   The phylogenetic tree of SARS-CoV-2 (nCov-19) viruses is reconstructed according to the similarity of genome sequences. The tree topology of Betacoronavirus is remarkably consistent with biologist's systematics. Because the tree construction contains enough information about virus mutants, it is suitable to study the evolutionary relationship between novel coronavirus mutants transmitted among humans. The emergences of 14 kinds of main mutants are studied and these strains can be classified as eight bifurcations of the phylogenetic tree.  It is found that there exist three types of virus mutations, namely, the mutation among sub-branches of the same branch, the off-root mutation and the root-oriented mutation between large branches of the tree. From the point of the relation between viral mutation and host selection we found that individuals with low immunity provide a special environment for the positive natural selection of virus evolution. It gives a mechanism to explain why large mutations between two distant branches generally occur in the nCov-19 phylogenetic tree. The finding is helpful to formulate strategies to control the spread of COVID-19.

**Introduction**

The rapidly expanding genetic diversity of SARS-CoV-2 (nCov-19) viruses provides a great deal of information about virus mutation. GISAID introduced a nomenclature system for major clades based on marker mutations [1]. Then, GISAID clades are augmented with more detailed lineages assigned by the Phylogenetic Assignment of Named Global Outbreak Lineages (PANGO lineage) tool [2]. Recently on 31-May-2021 World Health Organization (WHO) announced that the Greek alphabet will be used to refer to the main COVID-19 mutant virus in the future [3]. Now we found that more than ten variants – Alpha, Beta, Gamma, Delta, Kappa, Epsilon, Theta, Iota, Zeta, Lambda, Omicron etc., have been widely spread among humans. What is their evolutionary relationship?

The nCov-19 viruses spread in human is thought to belong to Sarbecovirus of Betacoronavirus that infers a possible bat origin [4-9]. However, there should exist some intermediate hosts in animal between the bat origin and the virus spreading in human. Although recent works discovered that the genome of Malayan pangolin coronaviruses shows high similarity to nCov-19, little is known about the problem of who is the real intermediate host [10-13]. Let us put the problem aside for the time being and in this paper we will focus on how the nCov-19 mutates in human. It is hoped that this will enlighten the problem of tracing source of nCov-19 and searching for its intermediate host.

About the calculation method of phylogenetic analysis, starting from information theory with the emphasize of base correlation property of the genome sequence we proposed IC-PIC method and used it in deducing the phylogenetic tree of Betacoronavirus [14-16]. The careful analysis of early 2019-nCoV tree was given in [16]. However, limited to the virus data at that time we were unable to study the various mutants of novel coronavirus. Now we will use IC-PIC method in this paper to make phylogenetic analysis based on a wealth of virus mutation data to reveal the law of how novel coronavirus mutates to produce evolutionary bifurcations.

**Methods**

We suggested the average mutual information (AMI) and k-departed base correlation can be looked as the signature of a given genome sequence [17]. The average mutual information is called information correlation (IC) defined by

$$D_{k+2} = -2\sum_i p_i \log_2 p_i + \sum_{ij} p_{i(k)j} \log_2 p_{i(k)j}$$

and the k-departed base correlation is called partial information correlation (PIC) defined by

$$F_{i(k)j} = (p_{i(k)j} - p_i p_j)^2$$

where $p_i$ means the probability of base $i$ in the sequence and $p_{i(k)j}$ means the joint probability of base pair $ij$ departed by distance $k$ ($k=0,1,2,…$). In the following we shall study the SARS-CoV-2 phylogeny by using IC-PIC algorithm based on the above set of signatures of the genome sequence.

The nCov-19 viruses genomes used in our analyses were downloaded from the GISAID (https://gisaid.org) platform. The genome sequence is converted into an IC-PIC matrix with 17 rows (representing 1 IC for given $k$ and 16 PICs of different base correlation categories) and d columns

(representing the distance *k* between base pair, *k*=0,1 to d-1). The only parameter in the algorithm is the range of d, which is denoted as **K**. **K** is determined from the best-fit construction of tree. In general the deduced tree changes with **K** and attains stable at some large value.

The work is carried out on IC-PIC web server [18]. After uploading input data in Fasta format, setting the parameter **K**-value and choosing the Neighbor-Joining (NJ) option, the server will run the program and for each run of given d (d=1 to **K** stepping 1) deduce a phylogenetic tree. In the calculation the evolutionary distance of any two genomes is calculated by Euclidean distance between their respective IC-PIC 17Xd matrices. Then an unrooted NJ tree is generated. Finally, **K** phylogenetic trees are combined to generate a consensus tree. All these trees were constructed by using NEIGHBOR and CONSENSE program in the PHYLIP package [19]. The robustness of the tree topology was estimated by branch support.

**Results and Discussions**

The whole-genome-based phylogenetic trees for nCov-19 are deduced by use of IC-PIC method and given in Fig 1. To reconstruct the phylogenetic tree the sequence data of 150 viruses are used. The consensus tree is derived from 50 (**K**=50) trees based on IC-PIC matric. The robustness of the tree topology was estimated by branch support. The NJ consensus tree is drawn with Avian viruses as out-group. However, the comparison between **K**=50 and larger K shows that the deduced tree attains stable at **K**=50. The tree topology remains also unchanged under different choices of the out-group.

Betacoronavirus contains five subgenus, namely Sarbecovirus, Hibecovirus, Nobecovirus, Embecovirus and Merbecovirus. The phylogenetic tree of Betacoronavirus genus is shown in Fig. 1. From Fig 1 we found that the tree topology is remarkably consistent with biologist's systematics that Embecovirus, Merbecovirus, Nobecovirus and Hibecovirus bifurcated first and then SARS-CoV-1 and 2 related and nCov-19 formed their own clade respectively [9-10]. We confirmed current taxonomic classification of the nCov-19 as Sarbecovirus sub-genus and supported the assumption that the nCov-19 is of bat origin. Above results show that our IC-PIC method is reliable. Further, in order to carefully study the variation of the virus as it spreads among humans, we have included 115 human virus mutation data in the tree of 150 sequences. The schematic diagram Fig 2 is deduced from the tree topology of Fig 1 for 115 human nCov-19 viruses. Figure 2 combined with Fig 1 gives the details of human novel coronavirus mutation that provide new understanding on the rapidly expanding genetic diversity of nCov-19 viruses. About the nCov-19 evolution in humans the phylogenetic analysis gives:

1) **Emergence of mutants and their evolutionary bifurcation**. The phylogenetic tree is deduced from the closeness relationship of sequence characteristics. The deduced phylogenetic tree of 115 human nCov-19 viruses shows eight main bifurcations which constitutes eight major clades of mutants (Fig 2). Two clades Omicron and Lambda are located in the lowest bifurcation near the root of the phylogenetic tree. Three mutants Alpha, Eta, and Gamma constitute a major clade emerging from the bifurcation No 1 of the phylogenetic tree. Two mutants Kappa and Delta constitute next clade emerging from the bifurcation No 2 of the phylogenetic tree. Three mutants Beta, Iota and Theta constitute another clade emerging from the bifurcation No 3 of the tree. Then Zeta, GH(Epsilon), G as independent clades emerge from the bifurcation No 4 to 6 respectively. Finally, L S and V constitute the last clade emerging from the bifurcation No 7 of the tree.

2) **The spatio-temporal localization of mutants and their dynamical variation and evolutionary trajectory.** L, S, V are early amino acid mutants.

They are thought to have no important amino acid mutation effects. D614G mutation in spike protein was firstly reported in Italy in February 2020 that obviously increases the infectivity of the virus. The mutant is named G. Then, in May 2020, L452R mutation was reported in USA that has strong ability of immune escape. The mutant is named Epsilon or GH. From GH to Omicron there are 12 important new mutants given in Fig 2. Note that all these mutants appeared in a given area that is because the geographical isolation plays a role in species formation. Each of them appeared also in a given time interval of 2020 apart from two (Theta and Omicron) in 2021. The time and place of the emergence of all these mutants were recorded in detail. Therefore, we are able to depict the evolutionary trajectories of mutants in principle. For example, Alpha firstly occurred in UK in September 2020 and then quickly spread to other countries (Germany, France, Italy, USA, Thailand and India, etc.). On the other hand, both Kappa and Delta occurred firstly in India in October 2020. It was reported Alpha Kappa and Delta coexisted in India in the early stage. At the beginning the three competed, and then Delta got the upper hand, becoming the main strain in India. One can assume Kappa and Delta have common precursor and Alpha is the possible candidate of the precursor. Another example is: Zeta and Gamma occurred firstly in Brazil in April 2020 and November 2020 respectively. In the early stage Zeta as the main strain was reported. Then it was reported Zeta and Gamma mixed and finally Gamma was the main strain in Brazil. One can assume Zeta is the precursor of Gamma. Moreover, the mutant Lambda occurred in Peru in July 2020 and it was mixed with Gamma in the early stage. One can assume Lambda and Gamma have common precursor. Above examples show that there exist a complex evolutionary network of human nCov-19 viruses which should be carefully searched for.

    3) **Three kinds of nCov-19 viral mutation.** The nCov-19 viral

mutation happens in a not-too-long time of only one or two years in the following three types: 1, the mutation between sub-branches in the same large branch; 2, the mutation occurring between adjacent or non-adjacent large branches in the direction deviating from the root (called off-root mutation); and 3, the mutation between large branches in the direction towards the root of the tree (called root-oriented mutation). The possible mutations between Alpha, Eta and Gamma, between Kappa and Delta, and between Beta, Iota and Theta belong to the first category. The possible mutations of Alpha to Delta, etc. belong to the second category. The possible mutations of Zeta to Gamma, LS+VS to G and G to GH (Epsilon), the emergence of Lambda in Peru in July 2020 and the emergence of Omicron in South Africa in November 2021, etc. belong to the third category. It is found most mutations belong to the third category.

    4) **Mechanism of virus mutation and natural selection**. Looking at the problem from the general point of view of evolutionary theory, why the nCov-19 mutation can occur between two distant branches and most in the direction towards the root? The evolution of virus is closely related to its host. The IC-PIC tree of viruses is constructed according to the sequence similarity. The sequences located on the same branch have stronger similarity and the mutation between them should have higher probability. Therefore the new species should gradually emerge at the top of the tree branches at given host. This is the usual pattern of virus evolution before it sneaked in humans. In fact, the source host of the virus is bat. The selection pressure of the virus in bats is very small. Bats had evolved a tolerant immune system that allows nearly all kinds of viruses replicate in their bodies [20]. On the other hand, although little is known about the intermediate hosts, if one assumes that the selection pressure felt by the viruses in the intermediate hosts is not great in most cases, then the viruses can adapt to the environment with fewer mutations generally.

The above examples are consistent with the construction of phylogenetic tree Fig 1 and do not violate the usual pattern of virus evolution. However, as the virus emerging into humans the human immune system is a special environment for viruses. Because the mutation is essentially at random, only those large virus mutations (for example, mutations between distant branches) are likely to include immune escape mutations and adapt to the environment. In other words, the immune escape mutation of nCov-19 needs external selection and the selection comes from incomplete immunity of the host (weak immunity to the virus). Therefore, from the relation between mutation and selection we infer that the immune escape mutations to be really selected can only happen in hosts with incomplete immunity[21]. The point that individuals with low immunity provide a special environment for virus evolution has been tested in experiments. As early as in the spread of Alpha in UK the nCov-19 evolution during treatment of chronic infection was studied [22]. It was observed that the mutant of the virus in the patient showed a reduced susceptibility to neutralizing antibodies and a higher level of viral infectivity. The recent emergence of Omicron in South Africa provides another evidence of virus fast evolution in people with low immunity. South Africa has many people infected with HIV which suppresses the immune system who were more suitable for nCov-19 evolution [23]. The above examples explained why the large mutations between two distant branches can be selected and the corresponding SARS-CoV-2 mutants do occur in experiments. In the meantime, since the root-oriented mutation has more possibilities than the off-root mutation in statistics, most observed nCov-19 mutations belong to the root-oriented category.

**Summary**

The emergence of nCov-19 variants that posed an increased risk to global public health prompted the global monitoring and research on their nucleotide mutations. To assist with public discussions of variants evolution, this paper summarizes various mutations that have taken place in the spread of nCov-19 in humans from a fundamental point of view, the view of phylogeny. First, the IC-PIC method is used to reconstruct the phylogenetic tree of Betacoronavirus and the tree topology is remarkably consistent with biologist's systematics. It gives supplementary evidence of the reliability of this method. Second, based on the phylogenetic tree of nCov-19 in humans and the spatio-temporal record of the virus mutations we found three mutation categories existing in the rapidly expanding genetic diversity of nCov-19, namely, the mutation among sub-branches, the off-root mutation and the root-oriented mutation between large branches of the tree, and the most mutations belong to the third category. Third, the relation of virus mutation and selection in humans is discussed. We found the individuals with low immunity provide a special environment for virus evolution. Since the condition of selection for virus evolution in humans might be very different from the usual pattern , large mutations between two distant branches and even the root-oriented mutations can occur.

Above conclusions are helpful to formulate COVID-19's prevention and control strategy. The nCov-19 mutations occurred mainly in 2020, but it still happened in 2021. There is no evidence that COVID-19 pandemic is coming to an end at Omicron. Oppositely, the longer the disease spreads, the more likely the virus is to mutate. Since the individuals with low immunity provide the positive selection to virus mutation, to minimize the mutation of the virus the international society should pay special attention to the health of these people. What is more noteworthy is that asymptomatic infected persons may have incomplete immunity [24] and they need vaccinated to strengthen their immunity.

## Acknowledgements

We gratefully acknowledge the authors and originating and submitting laboratories of the sequences from GISAID's EpiFlu(TM) Database on which this research is based.

# Figure caption

**Figure 1** - **The consensus tree (K=50) of nCov-19 with Gammacoronaviruses (avian viruses) as out-group**

　　The consensus tree of 150 nCov-19 is derived from 80 NJ trees based on IC-PIC matric, with Turkey-CoV and Avian infectious bronchitis virus as out-group. The tree has not been drawn to scale. The robustness of the tree topology was estimated by branch support. The comments on the left are arranged in the order in which the virus appears in the figure and the number given in parenthesis indicates the number of viruses used in the analysis. HCov-19 (115) means 115 viruses of human nCov-19, SARS-CoV-2 related (6) means 6 SARS-CoV-2 viruses , etc.

**Figure 2** - **Schematic diagram of virus mutation for nCov-19 transmitted in human**

　　The schematic diagram is deduced from the tree topology of Fig 1 for 115 human nCov-19 viruses. The number written at the node indicates the number of bifurcations. The number given in parenthesis indicates the number of viruses that the branch contains.

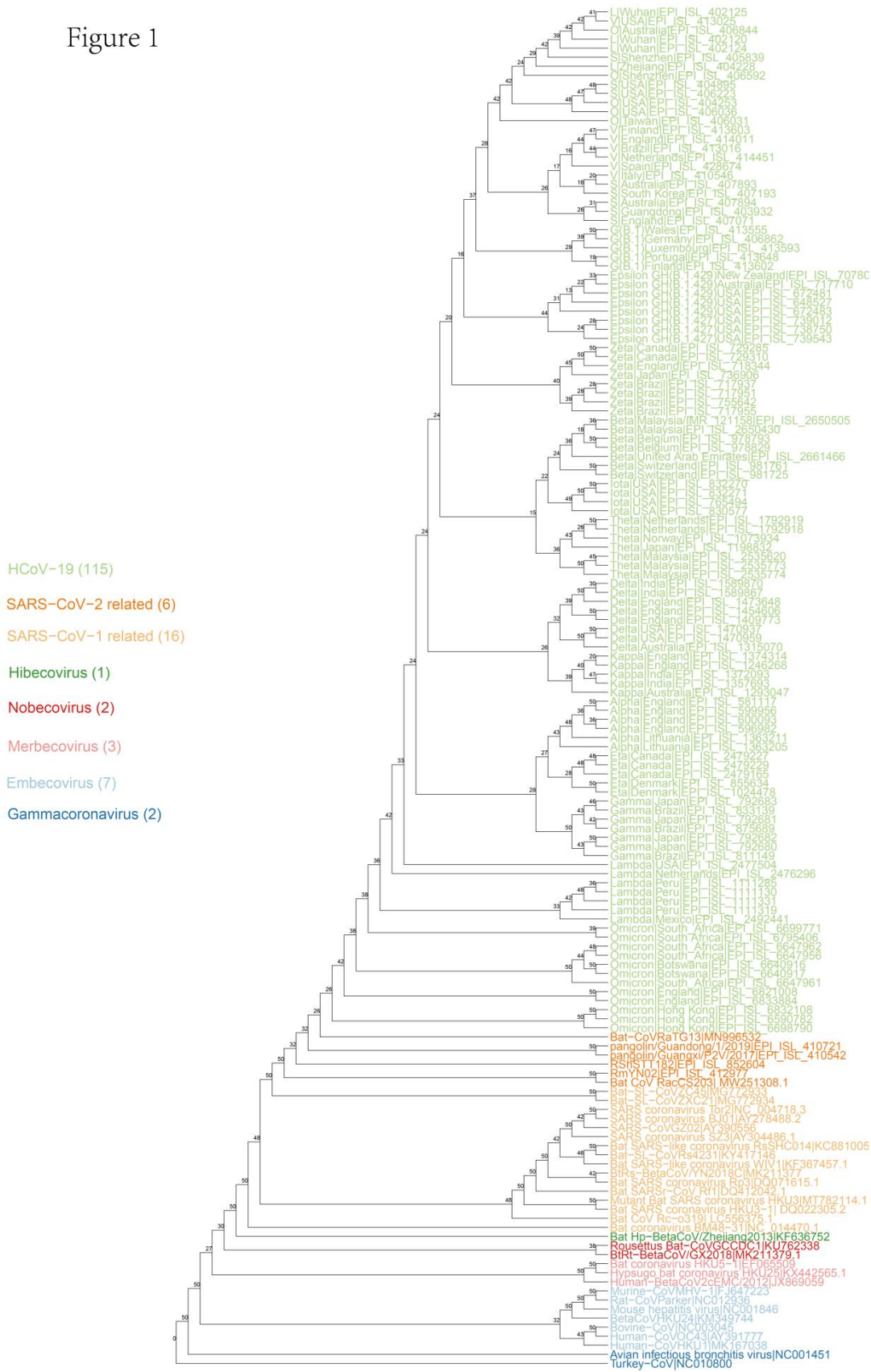

Figure 1

Figure 2

```
                                                    ┌── LS(13)
                                                ┌─8─┤
                                            ┌─7─┤   └── VS(11)
                                        ┌─6─┤   └────── G(5)
                                    ┌─5─┤   └────────── GH(8)
                                ┌─4─┤   └────────────── Zeta(8)
                                │   │        ┌── Beta(7)
                                │   │   ┌────┤
                            ┌─3─┤   └───┤    └── Iota(4)
                            │   │       └─────── Theta(7)
                        ┌─2─┤   │           ┌── Delta(8)
                        │   │   └───────────┤
                        │   │               └── Kappa(5)
                    ┌─1─┤   │           ┌── Alpha(6)
                    │   │   │       ┌───┤
                    │   │   └───────┤   └── Eta(5)
                ┌─0─┤   │           └────── Gamma(7)
                │   │   └────────────────── Lambda(7)
                │   └────────────────────── 
                └────────────────────────── Omicron(12)
```